\documentclass[10pt,a4paper,titlepage]{article}
\usepackage[utf8]{inputenc}
\usepackage{subfigure}
\usepackage{caption}
\usepackage{amsmath, amsfonts, amssymb}
\usepackage{graphicx, booktabs, xcolor}

\usepackage[left=2cm,right=2cm,top=3cm,bottom=3cm]{geometry}
\usepackage{authblk}
\usepackage[round]{natbib} 
\usepackage{multirow}
\usepackage{hyperref}
\usepackage{lscape}
\usepackage{color,soul}

\usepackage{algorithm}
\usepackage{algpseudocode}

\date{}
\title{A Study on the Optimal Design of Isothermal Experiments in Predictive Microbiology}

\author[1]{Alba Muñoz del Río} %
\author[1]{Víctor Casero-Alonso}
\author[1]{Mariano Amo-Salas}
\affil[1~]{Department of Mathematics, Institute of Mathematics Applied to Science and Engineering, University of Castilla-la Mancha}

\begin{document}
\begin{minipage}[h]{0.92\textwidth}
    \maketitle
    \begin{abstract}
    	This study addresses from the Optimal Experimental Design perspective the use of the isothermal experimentation procedure to precisely estimate the parameters defining models used in predictive microbiology. Starting from a case study set out in the literature, and taking the Baranyi model as the primary model, and the Ratkowsky square-root model as the secondary, $D$- and $c$-optimal designs are provided for isothermal experiments, taking the temperature both as a value fixed by the experimenter and as a variable to be designed. The designs calculated show that those commonly used in practice are not efficient enough to estimate the parameters of the secondary model, leading to greater uncertainty in the predictions made via these models. Finally, an analysis is carried out to determine the effect on the efficiency of the possible reduction in the final experimental time.
	\end{abstract}
\end{minipage}

\section{Introduction}

In the field of food safety, predictive microbiology has become an essential tool \citep{mcmeekin2007predictive}, as it is responsible for developing the models which allow the behaviour of the microorganisms in food to be described. These models can be divided into two main categories, primary and secondary models \citep{whiting1995microbial}. The primary models describe the growth or inactivation of the microorganisms in static or constant environmental conditions, and therefore these models are expressed only as functions of time. In the secondary models, the kinetic parameters of the primary model are expressed as functions of environmental factors, such as temperature, pH  or water activity.  \cite{dolan2013parameter} provides a comparison of the two types of model, with regard to the methodology for estimating the parameters, showing the main advantages and drawbacks of each one. 

Experimentation is key to the construction of these models, since the quality of experiments is directly related to the precision with which model parameters are estimated. This is very important, as uncertainty in the estimations is reflected in the predictions made with these models \citep{Vilas2018}.
Faced with this problem, many studies have considered using techniques of Optimal Experimental Design (OED), to determine the best experimental points to obtain the maximum possible information, and thus more accurate estimations. The work of \cite{versyck1999introducing}, pioneers in this area, applies OED in an Arrhenius-type secondary inactivation model. In the literature is widely used the combination of the Baranyi model as the primary model, and the Ratkowsky square-root model as the secondary, as it is the model type that has been most thoroughly analysed from the OED perspective, in both the isothermal and non-isothermal cases \citep{bernaerts2000design, bernaerts2002optimal, grijspeerdt2005practical, longhi2017optimal, longhi2018optimal}. Nevertheless, there are also studies applying OED to other  models, such as the modified Gompertz model \citep{gil2014application} or the Bigelow, Mafart and Peleg models \citep{penalver2019guidelines}.

When carrying out these experiments, there are two different possibilities: isothermal or non-isothermal experiments. Although the design of the experiments has been studied for both possibilities, various authors indicate the former are more appropriate for estimating the kinetic parameters of the model, due to the constant conditions throughout the experiments \citep{Braissant2010, Gil2011}. In addition, they are less expensive and easier to perform and analyse. In contrast, non-isothermal experiments provide a more realistic view of microbial behaviour. It is important to note that, in both types of experiments, the observations are obtained from different units of the food product under study. This ensures the use of genuine replicates, meaning that each measurement is taken independently by repeating the same experiment under identical conditions, thus avoiding correlation between observations. The usual procedure found in the literature for modelling microbial behaviour is to carry out isothermal experiments for different temperatures and estimate the primary model parameters for each temperature. From the estimates of the kinetic parameter of the model for the different temperatures, the parameters of the secondary model are estimated. Once all the parameters have been fitted, the model is validated using non-isothermal experiments \citep{DaSilva2020, Tarlak2020, Juneja2021}. However, this procedure involves a significant problem from a statistical point of view, since it assumes the estimates of the kinetic parameter as observed values in order to perform the subsequent fitting of the parameters of the secondary model, ignoring the uncertainty of these estimates. It is clear that if, for each temperature, the estimates of the kinetic parameter of two experiments are equal, they will lead to the same estimates of the secondary model parameters regardless of the uncertainty in the first. However, if this inherent variability is taken into account, it is assumed that the lower the uncertainty in the estimates of the kinetic parameter, the lower the uncertainty in the estimates of the parameters of the secondary model, and the greater the confidence in the predictions made by the model.

\section{Methods}

The aim of this study is to analyse, from the perspective of OED, the different scenarios that can be considered when planning isothermal experiments, in order to accurately estimate the parameters of the primary and secondary models. In this study, it is considered the primary Baranyi model and as a secondary model the Ratkowsky square-root model, because they are the most widely used in the literature. These models are used with the data obtained in the growth experiments of 
\textit{Pseudomonas spp.} on button mushrooms (\textit{Agaricus bisporus}), described by \cite{Tarlak2020}. This study shows both the optimal designs for each of these scenarios and the comparison between them, allowing the loss of efficiency that occurs in each case, and thus the robustness of the designs obtained, to be determined.  

\subsection{Mathematical models}

Among the models considered in the predictive microbiology literature, the Baranyi model stands out, due to its specificity and the physical meaning of its parameters. The explicit expression of this primary model \citep{baranyi1994dynamic} is as follows:

\begin{equation}
    y(t) = y_0 + \mu_{max}F(t) - \ln\left(1 + \frac{e^{\mu_{max}F(t)} - 1}{e^{(y_{max} - y_0)}}\right),
    \label{eq:Baranyi}
\end{equation}
with

\begin{equation*}
    F(t)= t + \frac{1}{\mu_{max}}\ln\left(e^{-\mu_{max} t} + e^{-\mu_{max}\lambda} - e^{-\mu_{max} t - \mu_{max}\lambda}\right),
\end{equation*} 
where \( y(t) \) is the natural logarithm of cell concentration (CFU/mL) at time \( t \) (hours); \( y_0 \) and \( y_{max} \) are the natural logarithms of initial and asymptotic concentrations, respectively; \( \mu_{max} \) is the maximum growth rate and \( \lambda\) is the lag phase duration (h).

With the aim of considering the effect of the temperature on the experiments, the Ratkowsky square-root model (\cite{ratkowsky1982relationship}) is taken as the secondary model. This describes the influence of suboptimal growth temperatures on the maximum specific growth rate of microorganisms ($\mu_{max}$). This model can be expressed as:

\begin{equation}
    \sqrt{\mu_{max}(T(t))}=b(T(t)-T_{min}),
    \label{eq:Ratkowsky}
\end{equation}
where $T$ is the temperature (\textdegree C), $b$ is a regression coefficient and $T_{min}$ is the theoretical minimum growth temperature.

\subsection{Optimal Experimental Design Methodology}

Starting from the premise that the total number of possible observations in an experiment is $N$, \textit{exact design} of size $N$ means a sequence of $N$ points in $\mathcal{X}$, where $\mathcal{X}$ represents the set of observable points, also known as the design space, on which the controllable variables are evaluated. This design is denoted $x_1, \ldots, x_N$, where some of these points may coincide. If $N_x$ is the number of observations to be made at point $x$, we define it as the number of genuine replicates at that point, meaning independent repetitions of the experiment under identical conditions at $x$. The discrete measure associated with that design is expressed as follows:

\[\xi(x) = \frac{N_x}{N}, \quad x \in \mathcal{X}.\]

 This gives the definition of  \textit{approximate design}, understood as a discrete probability measure \( \xi \) on $\mathcal{X}$, with finite support at $n$ distinct points, known as support points.
	\begin{equation*}
		\xi = \begin{Bmatrix}
			x_1 & x_2 &... & x_n\\
			w_1 & w_2 &... & w_n
		\end{Bmatrix}\in\Xi, ~~~~ \sum_{i=1}^{n}w_i = 1,
	\end{equation*}
where $\xi(x_i)= w_i$ are the weights or proportions observed at each point $x_i$ and $\Xi$ is the set of all approximate designs in $ \mathcal{X} $. Approximate designs are widely used in the OED literature due to their good properties for optimality calculation and verification. However, they must be adapted to the number of observations, $N$, to be carried out, which means rounding and fitting the weights. The calculation of efficient exact designs is considered in studies such as \cite{pukelsheim1992efficient}. Therefore, in this paper, the designs are considered approximate. \\
In this study, the model considered is a non-linear regression model

	\begin{equation*}\label{eq:3.3}
		y(x) =\eta(x, \theta)+ \varepsilon, ~~~~~~~\varepsilon\sim N(0, \sigma^2), ~~~~ x\in\mathcal{X}
	\end{equation*}
where $y$ is the response variable, $x$ the vector of designable or controllable variables taking values in $\mathcal{X}$, $\eta$ the regression function and $\theta$ is the vector of unknown parameters of the model. It is assumed that the error, $\varepsilon$, follows a normal distribution with mean zero and constant variance $\sigma^2$.\\
One of the main tools in this field is the Fisher Information Matrix (FIM) \citep{atkinson2007optimum}. By definition, the inverse of the FIM is asymptotically proportional to the variance-covariance matrix of the estimators of the model parameters. Therefore, optimising the estimation of the parameters of a model is equivalent to optimising the inverse of the FIM. Assuming normality in the observations, the FIM can be defined as
	\begin{equation*}
		M(\xi) = \sum_{x\in\mathcal{X}}f(x)f^T(x)\xi(x),
	\end{equation*}
where $f(x)=(f_1(x),...,f_k(x))^T$ and each $f_i(x)=\left.\frac{\partial \eta(x,\theta)}{\partial \theta_i}\right| _{\theta^{(0)}}$ corresponds to the partial derivative of the model with respect to the $i$-th parameter evaluated in $\theta^{(0)}=(\theta_1^{(0)},...,\theta_k^{(0)})^T$ which are the initial values of the parameters, called nominal values. The need for these nominal values leads to obtaining locally optimal designs, since it depends on the choice of these values.\\
Different optimality criteria arise from the varied ways in which the FIM can be optimised. Each of these criteria is defined via a criterion function, $\Phi$, which is convex and lower  bounded, and defined in $\mathcal{M}$, the set of all information matrices. The design $\xi^\star$ which minimises $\Phi[M(\xi^\star)]$ is called $\Phi$-optimal. This paper uses the most popular criterion in OED theory, the criterion of $D$-optimality. This is equivalent to minimising the volume of the confidence ellipsoid of the model parameter estimates, and so in practice it optimises the joint estimate of all the model parameters. Its criterion function is given by $\Phi_D[M(\xi)]=|M^{-1}(\xi)|$ when $|M(\xi)|\neq 0$\ and 0 otherwise. Usually the determinant of the FIM is maximised, given the property that the determinant of the inverse of a matrix is the inverse of its determinant.\\
Furthermore, the criterion of $c$-optimality is also considered, which allows estimating a linear combination of the parameters $c^T\theta$ with minimum variance, where $c$ is a known vector of constants. Its criterion function is given by $\Phi_c[M(\xi)]=c^TM^{-1}(\xi)c$. 

By means of the General Equivalence Theorem (GET) \citep{kiefer1960equivalence, white1973extension}, it is possible to verify the $\Phi$-optimality of a design for those criteria based on the FIM, using the sensitivity function $\phi(x, \xi)$. A design is optimal when $\phi(x,\xi^\star)\geq 0$ for all $x \in \mathcal{X}$ and $\phi(x,\xi^\star)=0$ at the support points of $\xi^\star$. In the case of the $D$-optimality criterion, the sensitivity function is expressed as $\phi(x, \xi) = k-d(x, \xi)$, where $d(x, \xi) = f^T(x)M^{-1}(\xi)f(x)$ is the generalised variance function, and $k$ represents the number of parameters of the model. For the $c$-optimality criterion, it is defined as $\phi(x,\xi) = c^TM^{-1}(\xi)c-(f^T(x)M^{-1}(\xi)c)^2 $ \citep{lopez2023optimal}.
For the calculation of the optimal designs, it has been implemented in Python the Wynn-Fedorov algorithm, which is based on the GET \citep{wynn1972results, fedorov2013theory}.

\subsection{Efficiency and sensitivity analysis}

The $\Phi$-efficiency of a design $\xi \in \Xi$ is a measurement that evaluates the quality of the design $\xi$ as compared to the $\Phi$-optimal design, $\xi^\star$. This efficiency takes values between 0 and 1, and is defined by the following expression:
\begin{equation}
    \Phi\text{-Eff}(\xi|\xi^\star) = \frac{\Phi[M(\xi^\star)]}{\Phi[M(\xi)]}.
    \label{eq:eficiencia}
\end{equation}
It is common to calculate the efficiency of a design with respect to the optimal design, although it is also possible to determine relative efficiencies between different designs. In the case of relative efficiencies, the set of possible values the efficiency can take is $[0, \infty)$.\\
The Atwood bound allows a minimum bound to be obtained for the efficiency of a design. It is used here to guarantee a minimum efficiency in cases where the Wynn-Fedorov algorithm has computational problems with convergence to the optimal design, as can occur in $c$-optimal designs, usually singular. This bound can be expressed as

\begin{equation*}
    \Phi\text{-Eff}(\xi|\xi^\star) \geq \frac{Tr[M(\xi)\nabla\Phi(\xi)]}{\max_{x\in\mathcal{X}} d(x,\xi)},
\end{equation*}
where $Tr$ is the trace and $\nabla\Phi(\xi)$ is the gradient of the criterion function $\Phi$.\\
Given the uncertainty in the choice of nominal values for parameters, it is useful to perform a sensitivity analysis for the design with respect to these nominal values. This analysis studies the efficiency of the design when the chosen nominal value is not the true value of the parameter. This is done by varying the nominal value within a range considered of interest and calculating the efficiency of the design in each case. When the variations in efficiency are significant, it is said that the design is sensitive to uncertainty in that parameter, while in the opposite case it is said that the design is robust.

\subsection{Dataset}

As a reference to explain a standard procedure followed in predictive microbiology for the estimation of the parameters of primary and secondary models, the experiment carried out by \cite{Tarlak2020} is taken as a case study. 

This experiment first considers four isothermal experiments at temperatures of 4, 12, 20 and 28\textdegree C (shown in Table \ref{tab:diseños Tarlak}). The results of these experiments give estimates for the kinetic parameters $y_0, y_{max}, \mu_{max}$ and $ \lambda$ of the primary model expressed by Equation \eqref{eq:Baranyi} for each temperature, as shown in Table \ref{tab:estimaciones Tarlak}. Next, the estimate for the parameter $\mu_{max}$, associated with each temperature, is taken as the observed value, to obtain the estimates for the parameters of the secondary model ($b=0.0099\pm 4.95\times 10^{-4}$ and $T_{min}=-17.5\pm 1.7$\textdegree C) defined by Equation \eqref{eq:Ratkowsky}. Finally, these models are validated by expressing them in the form of differential equations using non-isothermal experiments.

Based on this procedure, different experimentation strategies addressed through OED are provided for each of the possible scenarios.

\begin{table}[h!]
\centering
\begin{tabular}{c|cccccccc}
\multicolumn{1}{l|}{T (\textdegree C)} & \multicolumn{8}{c}{Sampling points (hours)}                        \\ \hline
$4$                         &  0 & 24 & 60 & 96 & 144 & 192 & 240 &     \\ 
$12$                        &  0 & 24 & 48 & 72 & 100 & 144 & 192 &     \\ 
$20$                        &  0 & 12 & 24 & 36 & 48  & 72  & 96  & 120 \\ 
$28$                        &  0 & 12 & 24 & 36 & 48  & 60  & 72  & 84 
\end{tabular}
\caption{Isothermal designs in the experiment carried out by \cite{Tarlak2020}.}
\label{tab:diseños Tarlak}
\end{table}

\begin{table}[h!]
\centering
\begin{tabular}{c|cccc}
\multicolumn{1}{l|}{T (\textdegree C)} & $y_0$  (log CFU/g)         & $y_{max}$   (log CFU/g)     & $\mu_{max}$ (h$^{-1}$)      & $\lambda$  (h)     \\ \hline
$4$                         & $7.08 \pm 0.06$ & $8.65 \pm 0.05$  & $0.043 \pm 0.007$ & $44.1 \pm 10.4$ \\ 
$12$                        & $7.08 \pm 0.06$ & $10.14 \pm 0.05$ & $0.091 \pm 0.006$ & $33.2 \pm 3.6$  \\ 
$20$                        & $7.06 \pm 0.12$ & $10.71 \pm 0.13$ & $0.134 \pm 0.015$ & $23.5 \pm 4.7$  \\ 
$28$                        & $7.05 \pm 0.07$ & $10.64 \pm 0.07$ & $0.202 \pm 0.013$ & $20 \pm 1.9$   
\end{tabular}
\caption{Estimation of the parameters for the Baranyi model in \cite{Tarlak2020}.}
\label{tab:estimaciones Tarlak}
\end{table}

\section{Results}

\subsection{Optimal designs in isothermal experiments}

\subsubsection{$D$-optimal designs}\label{Sec:3.1.1}
\begin{figure}[h!]
    \centering
    \subfigure[T=4\textdegree C]{\includegraphics[width=0.4\textwidth]{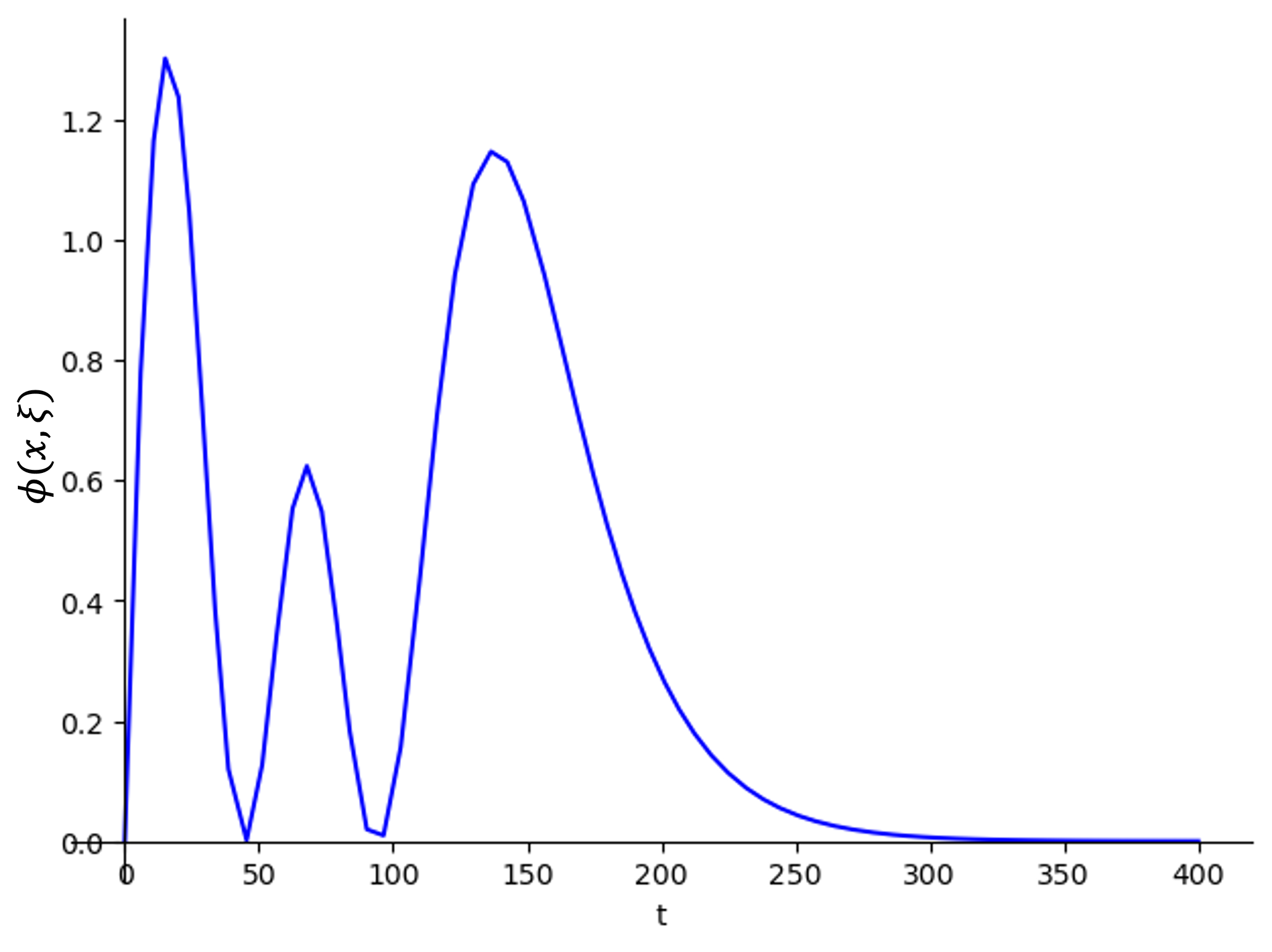}}
    \subfigure[T=12\textdegree C]{\includegraphics[width=0.4\textwidth]{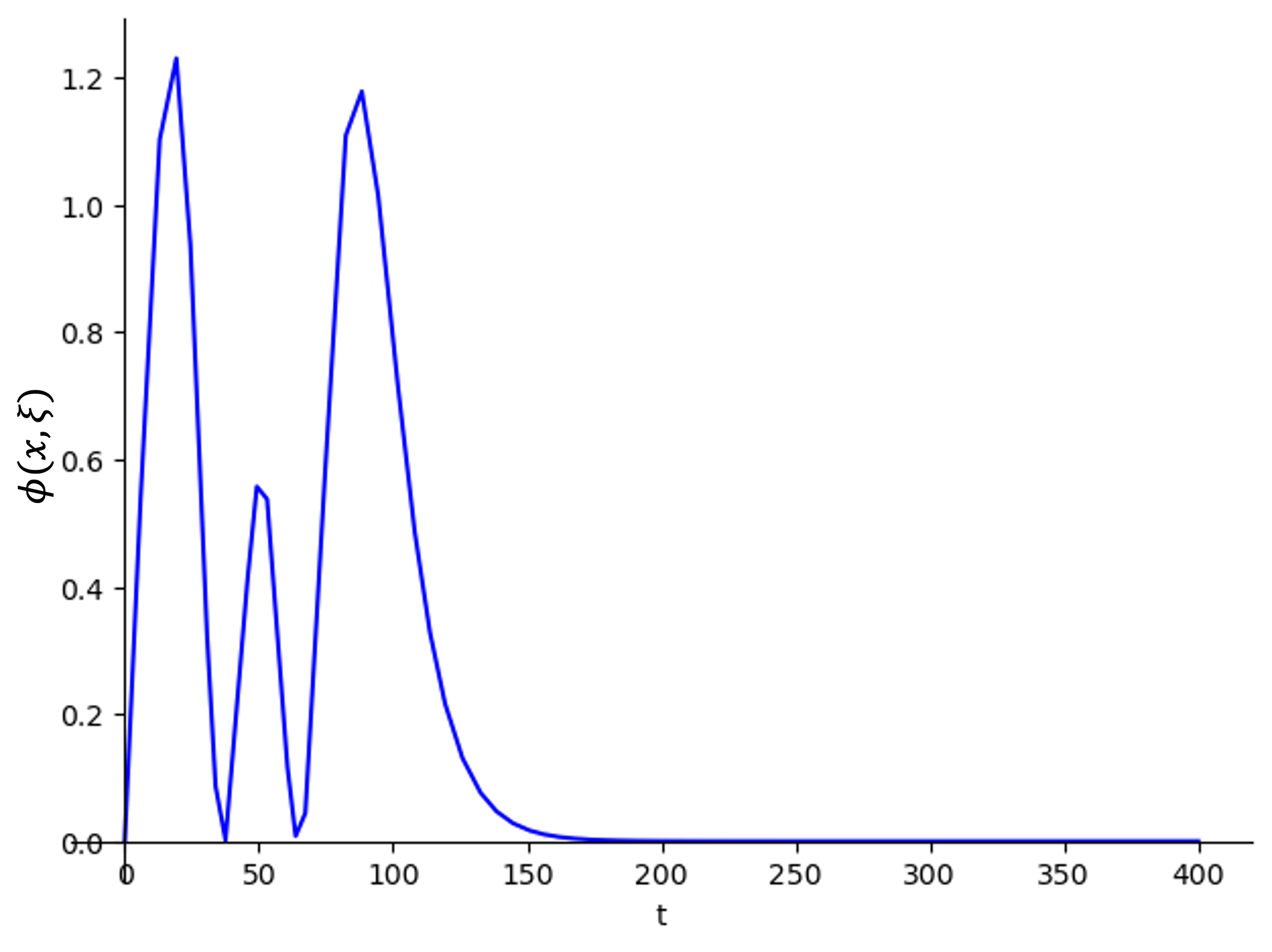}}\\
    \subfigure[T=20\textdegree C]{\includegraphics[width=0.4\textwidth]{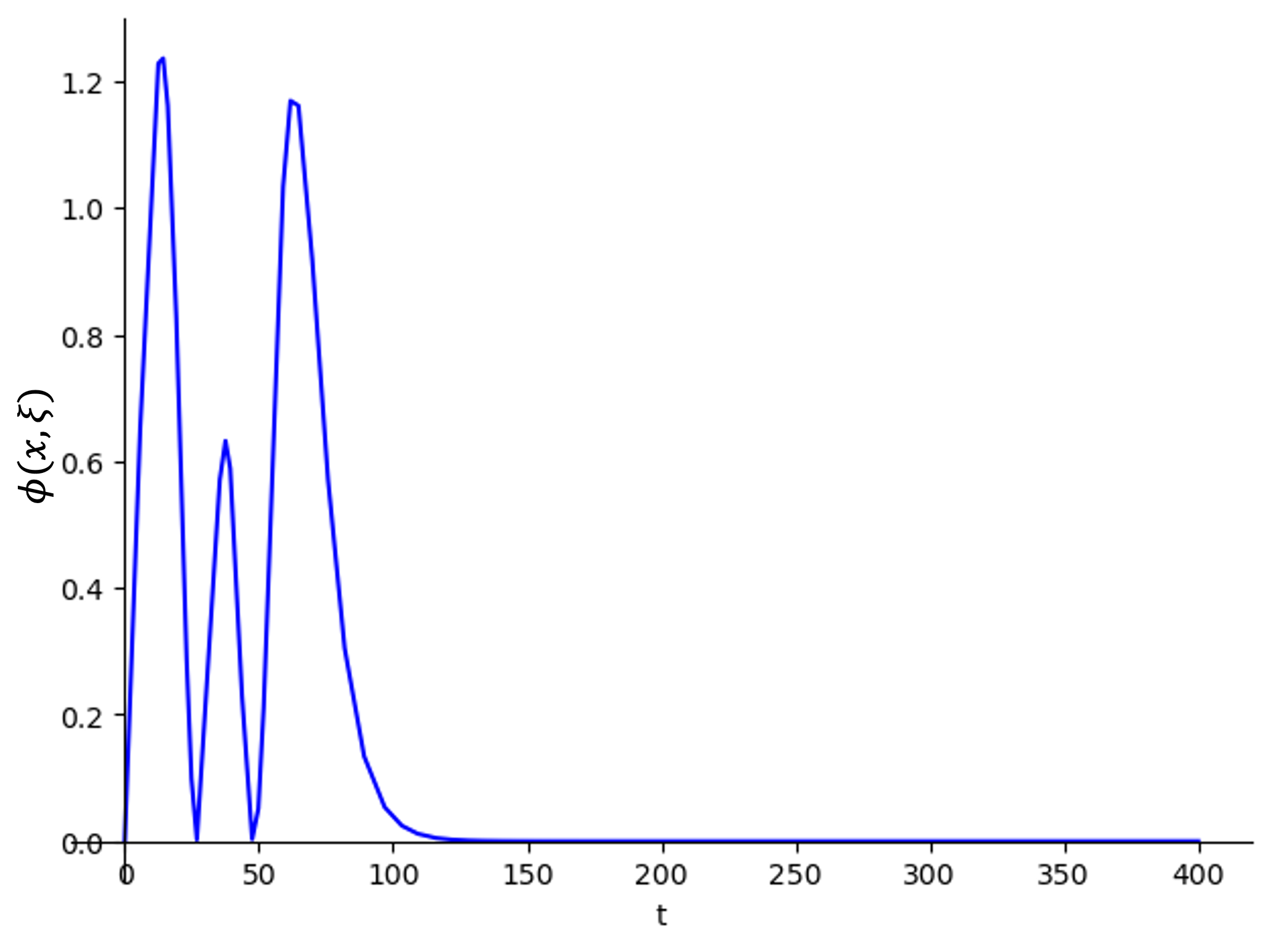}}
    \subfigure[T=28\textdegree C]{\includegraphics[width=0.4\textwidth]{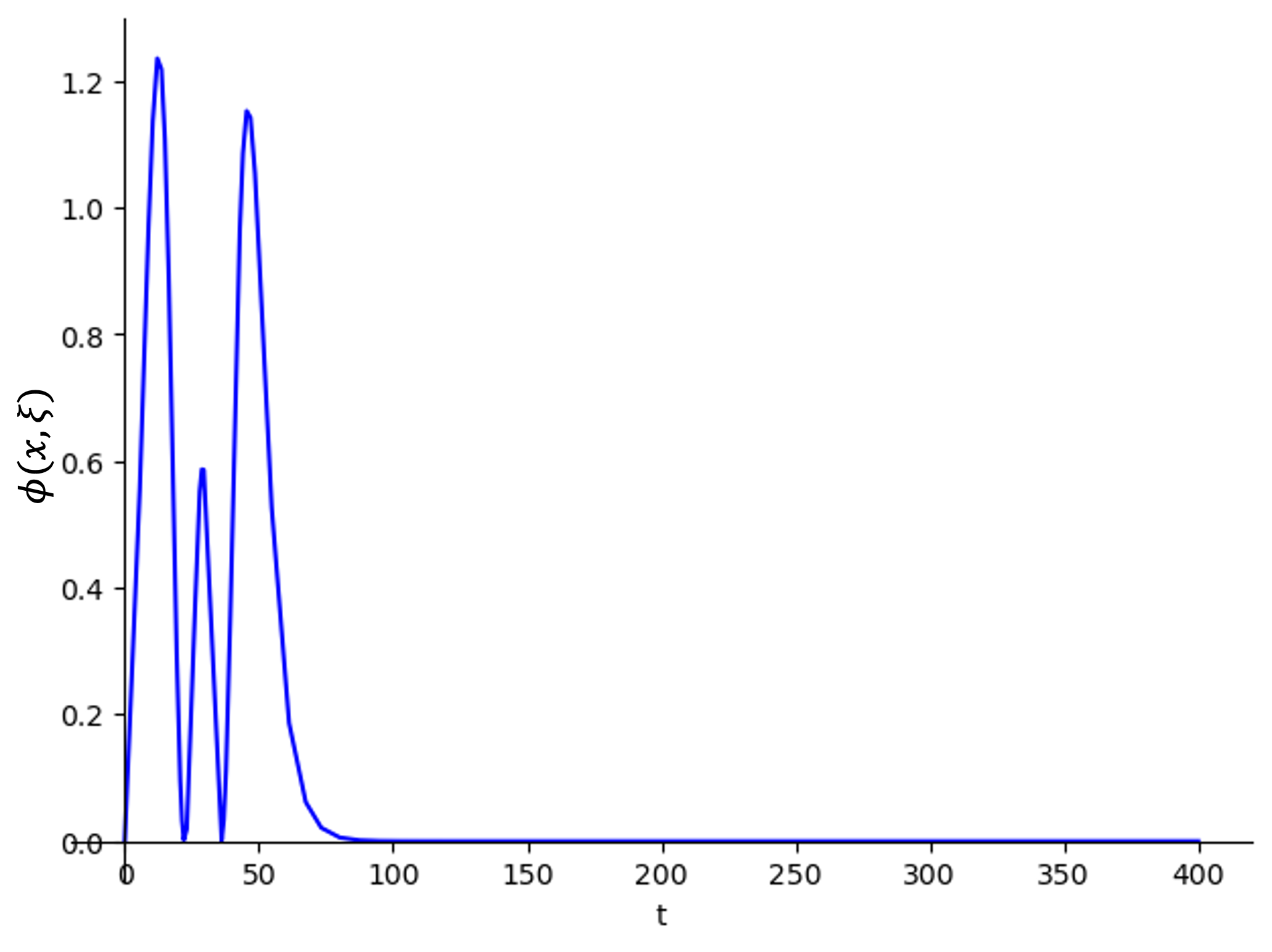}}
    \caption{Sensitivity function for $D$-optimal designs in Table \ref{tab:d-opt-isotermicos}.} \label{fig:4.5}
\label{fig:fun-var-d-opt}
\end{figure}

Firstly, to estimate precisely all the model parameters ($y_0, y_{max}, \mu_{max}, \lambda$), $D$-optimal designs are given for each temperature proposed by \cite{Tarlak2020}. 
The design space for time is considered from the beginning of the experiment (0h) until the stationary phase of the model is reached, at which population growth no longer occurs. The parameter estimates provided by the authors for each temperature are considered as nominal values (Table \ref{tab:estimaciones Tarlak}). Table \ref{tab:d-opt-isotermicos} shows the optimal designs obtained for each temperature, and the $D$-efficiency of the design proposed in the work of \cite{Tarlak2020} for each temperature. To calculate these efficiencies, their designs are considered approximate, with weight $1/N$ at each observation time. The $D$-optimal designs obtained are equally-weighted designs, with each support point corresponding to 25\% of the total observations. These efficiencies suggest that the precision of the joint estimation of the parameters is between 14\% and 20\% lower than that of the optimal designs. 
Figure \ref{fig:fun-var-d-opt} shows the sensitivity functions of the $D$-optimal designs. It is clear that the support points of the design are distributed similarly for the four cases, observing shorter distances between times the higher the temperature of the experiment. This behaviour is expected, since the higher the temperature, the faster the growth of microorganisms. 

\begin{table}[h!]
\centering
\begin{tabular}{c|c|c}
\multicolumn{1}{l}{T (\textdegree C)}        & \multicolumn{1}{c|}{$D$-optimal design}    & \multicolumn{1}{l}{$D$-eff} \\ \hline
\multicolumn{1}{c|}{$4$}  & $\begin{Bmatrix} 0 & 45   & 92   & 485 \\ 1/4 & 1/4 & 1/4 & 1/4 
\end{Bmatrix}$ & 0.83                      \\
\multicolumn{1}{c|}{$12$} & $\begin{Bmatrix} 0 & 37.2 & 65.2 & 280 \\ 1/4 & 1/4 & 1/4 & 1/4 
\end{Bmatrix}$  & 0.82                      \\ 
\multicolumn{1}{c|}{$20$} &$ \begin{Bmatrix} 0 & 27.5 & 48   & 185 \\ 1/4 & 1/4 & 1/4 & 1/4 
\end{Bmatrix}$  & 0.86                      \\ 
\multicolumn{1}{c|}{$28$} &$\begin{Bmatrix} 0 & 22.1 & 35.8 & 122 \\ 1/4 & 1/4 & 1/4 & 1/4 
\end{Bmatrix} $  & 0.8                      
\end{tabular}
\caption{Equally-weighted $D$-optimal designs for estimating $y_0$, $y_{max}$, $\mu_{max}$, and $\lambda$, and $D$-efficiencies of designs of Table \ref{tab:diseños Tarlak} with respect to $D$-optimal designs.}
\label{tab:d-opt-isotermicos}
\end{table}

\subsubsection{$c$-optimal designs for $\mu_{max}$}\label{Sec:3.1.2}
Taking the $\mu_{max}$ parameter estimate as the observed value for estimating the secondary model parameters leads to ignoring its uncertainty in the second estimate. In this case, the best approach is to use an estimate of $\mu_{max}$ with minimum variance, and so $c$-optimality should be considered, which provides designs to estimate the parameter $\mu_{max}$ with minimum variance for each temperature. Table \ref{tab:c-opt mumax} shows these $c$-optimal designs for each temperature. It should be noted that in this case, equally-weighted designs are no longer obtained, but rather certain support points have a greater weight than others in the design. Thus, the central points of the design account for between 65\% and 70\% of the observations to be made. This is related to the physical meaning of the kinetic parameter $\mu_{max}$, as it defines the slope of the growth curve in the central phase, the exponential phase. A certain similarity can also be seen between the points and the weights of the various designs, and the phenomenon already mentioned for the $D$-optimal designs, whereby the higher the temperature the shorter the time between support points, is seen again. These designs allow the parameters of the secondary model to be estimated based on the values of $\mu_{max}$ with minimal uncertainty. The third column of Table \ref{tab:c-opt mumax} gives the $c$-efficiency of the designs proposed by the authors (Table \ref{tab:diseños Tarlak}) with respect to the $c$-optimal design; that is, the goodness measure of how well they estimate the parameter $\mu_{max}$ with respect to the $c$-optimal design. These $c$-efficiencies, below 50\% in all cases, indicate that the parameter estimation of these designs entails 50\% more variability than that obtained with the $c$-optimal design.
\par
It is clear that these designs do not provide accurate individual estimations of the parameter $\mu_{max}$. The last column of Table \ref{tab:c-opt mumax} shows the $c$-efficiency of the $D$-optimal designs obtained in Section \ref{Sec:3.1.1}, confirming that, even using these designs, the estimation of parameter $\mu_{max}$ is considerably more accurate than with those previously mentioned.

\begin{table}[h!]
\centering
\begin{tabular}{c|c|c|c}
\multicolumn{1}{l|}{T (\textdegree C)} & $c$-optimal design                                   & \multicolumn{1}{l|}{Tarlak $c$-eff} & $D$-optimal $c$-eff \\ \hline
$4$                                                   & $\begin{Bmatrix} 0     & 41    & 97.4  & 300 \\ 0.14  & 0.295 & 0.36  & 0.205
\end{Bmatrix} $    & 0.47                                & 0.82              \\
$12$                                                  & $\begin{Bmatrix} 0     & 34.6  & 67.34 & 200 \\ 0.145 & 0.33  & 0.345 & 0.165
\end{Bmatrix} $ & 0.44                                & 0.82              \\
$20$                                                  & $\begin{Bmatrix} 0     & 25.8  & 49.5  & 125 \\ 0.137 & 0.34  & 0.36  & 0.153
\end{Bmatrix} $   & 0.44                                & 0.8               \\
$28$                                                  & $\begin{Bmatrix} 0     & 21.15 & 36.97 & 100 \\ 0.145 & 0.337 & 0.342 & 0.151
\end{Bmatrix} $ & 0.42                                & 0.81             
\end{tabular}
\caption{$c$-optimal designs for estimating $\mu_{max}$.}
\label{tab:c-opt mumax}
\end{table}

\subsection{Optimal designs controlling temperature and time}

In the optimal designs calculated in the previous section, the temperatures considered are the ones fixed by the experimenters in the work of \cite{Tarlak2020}. It is possible, however, that the temperatures chosen in the experiments also influence the estimation of the model parameters. Therefore, to accurately estimate the parameters of the primary model (function of time) and the secondary model (function of temperature), the most suitable solution is to design over both variables, time and temperature. To do this, the design is calculated for the primary model (Equation  \eqref{eq:Baranyi}) in which the parameter $\mu_{max}$ is substituted for the secondary model (Equation \eqref{eq:Ratkowsky}). Calculating the $D$-optimal design for this extended model provides the design that allows accurate estimation of the parameters $y_0$, $y_{max}$, $\lambda$, $b$ and $T_{min}$, jointly. In this way, not only the times but also the temperatures at which the experiment should be carried out are obtained, thus eliminating the constraint of having to experiment at the fixed temperatures of 4, 12, 20 and 28\textdegree C. The design space considered for temperature is [4\textdegree C, 28\textdegree C].\\

\subsubsection{$D$-optimal design}\label{Sec:3.2.1}

\begin{figure}[h!]
    \centering
    \includegraphics[width=0.47\textwidth]{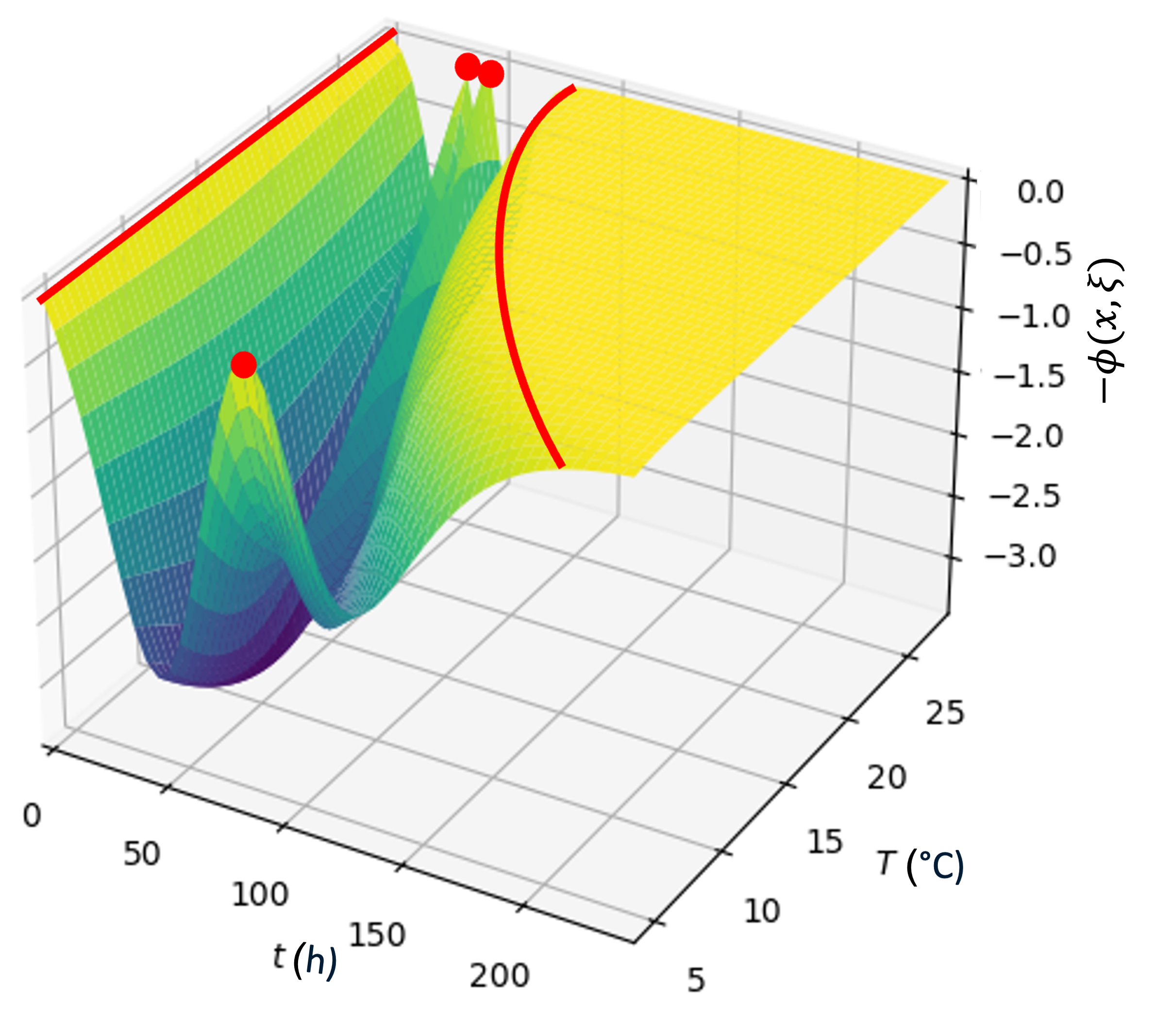}
    \caption{Isothermal $D$-optimal design in temperature ($T$) and time ($t$) for Tarlak.} 
    \label{fig:D-opt_T_t}
\end{figure}

In order to calculate the $D$-optimal isothermal design in temperature and time, we take as nominal values of the primary model parameters ($y_0$, $y_{max}$ and $\lambda$) the mean of the estimates provided by \cite{Tarlak2020}.  For the parameters of the secondary model, $b$ and $T_{min}$, the nominal values taken are the estimates provided in that work. 
Figure \ref{fig:D-opt_T_t} shows, for ease of visualisation, the opposite of the $D$-optimal design sensitivity function where the maxima, marked in red, correspond to the support points of the design.
This design leaves the temperature of the first point to the experimenter's choice, since the maximum is reached for time 0 at any temperature. 
In the same way, the last support point of the design could be freely chosen from those that make up the flat surface that extends to the end of the experimentation time, that is, the time until which it would be necessary to observe in order to make the final observation. 
Considering that the optimal cost option is to stop experimenting as soon as possible, the optimal lengths of the experiments are calculated for the full temperature range (4\textdegree C to 28\textdegree C) for which optimality is achieved in the design. The relationship between the temperature at which the experiment is to be performed and the final experimental time is then adjusted, using these values, via the function $t_f(T)=744.7-186.8\ln(T)$. This relationship also explains the final experimental time of the experiment for the isothermal $D$-optimal designs in Table \ref{tab:d-opt-isotermicos}. Thus, the $D$-optimal design obtained is \\

\begin{equation*}
     \xi^\star_T = \begin{Bmatrix}
        (T,0) & (4,87.88) & (28,31.17) & (28,43.75)  & (T,t_f(T)) \\
        0.2 & 0.2 & 0.2 & 0.2 & 0.2 
    \end{Bmatrix}.
\end{equation*}

Note that this design could be applied when performing only two isothermal experiments, one at 4\textdegree C and the other at 28\textdegree C. Since at 0h it would be possible to experiment at any temperature, either 4\textdegree C, or 28\textdegree C could be chosen. 
For the final experimental point, if the temperature is chosen to be 4\textdegree C, the final experimental time would be $t_f(4)=485.74$h, while if the decision is to experiment at 28\textdegree C, it would be $t_f(28)=122.24$h.
Thus, for example, if the cost of increasing the temperature were high, the choice could be made to carry out three independent observations at 4\textdegree C at $0$h, $87.88$h and $t_f(4)=485.74$h and two independent observations at 28\textdegree C, at $31.17$h and $43.75$h. While if the time cost is  higher, then observations could be made at 4\textdegree C at two time points, $0$h and $87.88$h, and at 28\textdegree C at three points $31.17$h, $43.75$h and $t_f(28)=122.24$h. Any of the above combinations maintains the optimality of the design, that is, it optimises precision in the joint estimation of all parameters.

For the purposes of making a comparison, the Tarlak design in two variables is considered to be the union of the four designs that they propose at each of the temperatures (Table \ref{tab:diseños Tarlak}). Thus, each experimental point has an associated weight of 1/30. The $D$-efficiency of this design with respect to $\xi^\star_T$ is 0.58. In the case of the joint $D$-optimal design, from the union of the four designs described in Table \ref{tab:d-opt-isotermicos}, its $D$-efficiency with respect to $\xi^\star_T$ is 0.65. The efficiencies indicate that the designs applied previously, with the imposition of the fixed temperatures, are rather limited when it comes to accurately estimating the parameters of both models together.

\subsubsection{Sensitivity analysis}

\begin{figure}[h!]
    \centering
    \subfigure[]{\includegraphics[width=0.32\textwidth]{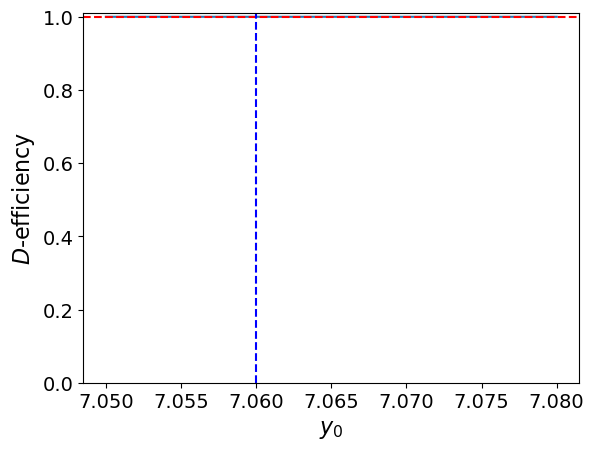}}\hfill
    \subfigure[]{\includegraphics[width=0.32\textwidth]{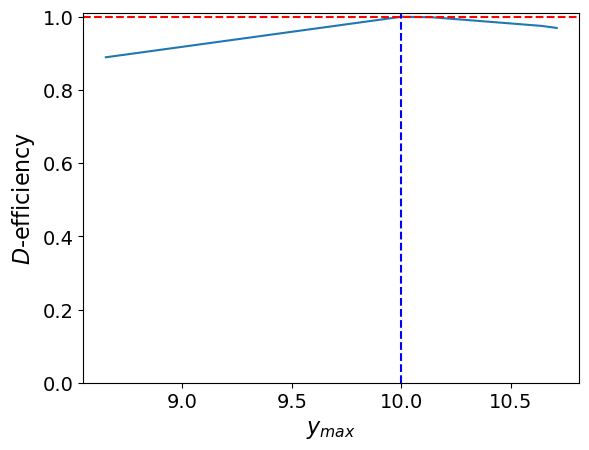}}\hfill
    \subfigure[]{\includegraphics[width=0.32\textwidth]{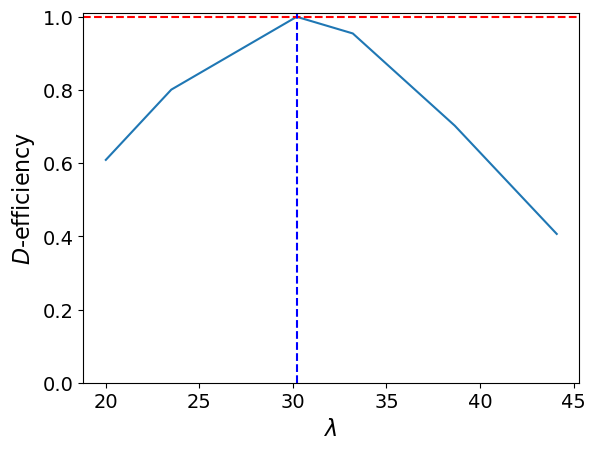}}\\
    \caption{Sensitivity analysis for the primary model parameters $y_0$ (a), $y_{max}$ (b) and $\lambda$ (c).} 
    \label{fig:sensib_analysis}
\end{figure}

Since the nominal values of the primary model parameters, $y_0$, $y_{max}$ and $\lambda$,  have been taken as the mean of the estimates of these parameters obtained by \cite{Tarlak2020} (Table \ref{tab:estimaciones Tarlak}), it is of interest to study the sensitivity of the $D$-optimal design $\xi^\star_T$ to the range of values of the estimates of these parameters. To carry out this analysis, the efficiency of the obtained $D$-optimal design, $\xi^\star_T$, is calculated with respect to the $D$-optimal design for another nominal value of this parameter using Equation \eqref{eq:eficiencia}. 
Figure \ref{fig:sensib_analysis} provides a graphical representation of the sensitivity analysis. The red line indicates the maximum achievable efficiency, which is equal to 1. For this reason, the optimal design achieves maximum efficiency for the initial nominal value (marked by the vertical blue dashed line). As shown in Fig. \ref{fig:sensib_analysis}, this design undergoes a greater loss of efficiency for variations in the nominal value of the parameter $\lambda$, and is more pronounced in the case of underestimation of this parameter, while it behaves robustly for the range of estimates of the parameters $y_0$ and $y_{max}$. 
As a solution to the loss of efficiency with respect to the uncertainty in the nominal values, \cite{del2024new} describe a methodology through which the optimal design is made robust.

\subsubsection{Analysis of the relationship between temperature and final experimental time}

\begin{figure}[h!]
    \centering
    \subfigure[Real values obtained]{\includegraphics[width=0.4\textwidth]{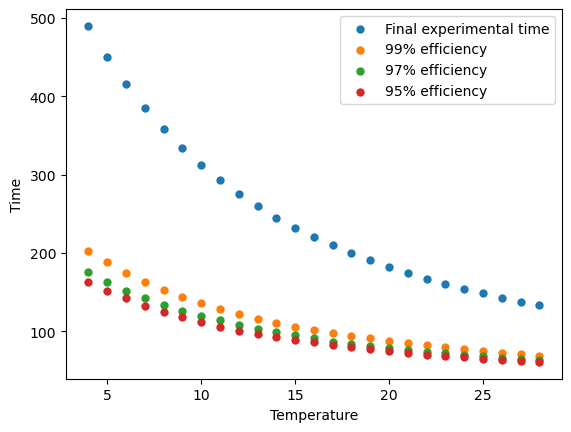}}
    \subfigure[Curve fitting]{\includegraphics[width=0.4\textwidth]{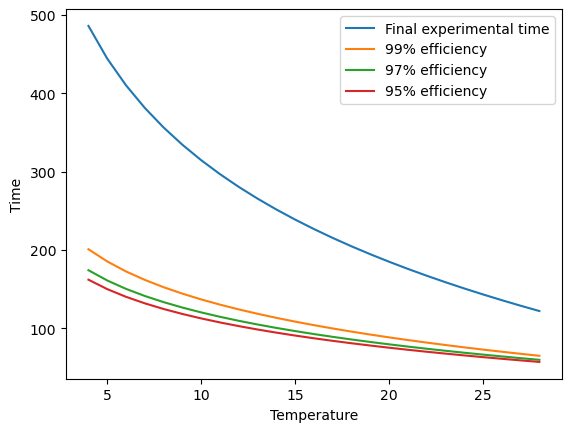}}\\
    \caption{Final time-saving curves of the experiment ensuring minimum efficiency.} 
    \label{fig:curvas ahorro}
\end{figure}

As shown in Section \ref{Sec:3.2.1}, $\xi^\star_T$ has a support point where the experiment time depends on the temperature at which the experiment is being carried out. As shown in the results, as the temperature increases, the  final experimental time decreases. Furthermore, this relationship between final experimental time and temperature has been adjusted using a logarithmic model. It seems logical that for cost-saving reasons the experimenter would therefore seek to shorten the experiment as much as possible. Therefore, based on the $\xi^\star_T$ design, an efficiency analysis was carried out to determine the final experimental times that lead to designs with high efficiency, and so can lead to greater time savings. Figure \ref{fig:curvas ahorro} (a) shows the behaviour of the relationship between temperature and final experimental time for both the $\xi^\star_T$ design and for designs with 99\%, 97\% and 95\% efficiency. It can be seen that assuming a 1\% efficiency loss already significantly reduces the final experimental times with a time saving of between 48\% and 58\%. Table \ref{tab:relaciones_temp_tiempo} also shows the adjustments to these relationships (represented in Figure \ref{fig:curvas ahorro} (b)) along with the RMSE value and the relative savings in experiment time. It should be noted that the best adjustments found after considering different options, such as polynomial and exponential functions, are shown.

\begin{table}[h!]
\centering
\begin{tabular}{c|c|c|c|c}
Minimum efficiency & $t_f(T)$            & RMSE & \begin{tabular}[c]{@{}c@{}}Minimum\\ saving\end{tabular} & \begin{tabular}[c]{@{}c@{}}Maximum\\ saving\end{tabular} \\ \hline
99\%              & $297.8-69.8\ln(T)$  & 1.55 & 48.5\%                                                  & 58.6\%                                                  \\
97\%              & $255.7-58.72\ln(T)$ & 1.62 & 52.9\%                                                  & 64\%                                                    \\
95\%              & $236.9-53.88\ln(T)$ & 2.01 & 54.5\%                                                  & 66.7\%                                                 
\end{tabular}
\caption{Relationship between temperature and final experimental time based on efficiency. }
\label{tab:relaciones_temp_tiempo}
\end{table}

\subsubsection{$c$-optimal design for $b$ and $T_{min}$}

Although the solution proposed in Section \ref{Sec:3.1.2} represents a significant improvement when adjusting the parameters of the secondary model, a design solely intended to estimate these parameters solves the problem of avoiding the uncertainty associated with the estimation of the parameter $\mu_{max}$ when it is taken as an observed value. The $c$-optimal design of both parameters simultaneously is proposed for this purpose

\begin{equation*}
    \xi^\star_{b,T_{min}} = \begin{Bmatrix}
        (T,0) & (4,111.5) & (26.3,60.7)   & (28, 44.4) & (28,160)  \\
        0.188 & 0.287     & 0.32            & 0.196      & 0.0038 
    \end{Bmatrix}.
\end{equation*}

The $c$-efficiency of the $D$-optimal design, $\xi^\star_T$, with respect to $\xi^\star_{b,T_{min}}$ is 0.015, indicating very poor precision in estimating the secondary model parameters with the design intended for accurate estimation of all parameters together.
The option of calculating the $c$-optimal for $b$ and $T_{min}$ separately was also explored. In the latter case, the design is the same as the joint $c$-optimal design for $b$ and $T_{min}$. In the case of the $c$-optimal design for estimating parameter $b$, its $c$-efficiency with respect to the $c$-optimal design for both parameters, $\xi^\star_{b,T_{min}}$, is 0.41, so it does not represent a direct solution to the problem.

\section{Discussion and Conclusions}

This study addresses the experimental procedure used in predictive microbiology for estimating model parameters from the OED perspective. The Baranyi model is considered as the primary model, and the Ratkowsky square-root model as the secondary model applied to the case study set out in \cite{Tarlak2020}. 
Firstly, considering the isothermal experiments with the temperatures fixed and the objective being to estimate all the parameters of the primary model, the $D$-optimal design is provided for each temperature. 
The $D$-efficiency (Table \ref{tab:d-opt-isotermicos}) of the designs used in practice to estimate these parameters shows a loss of efficiency of around 20\% compared to the optimal ones, which results in lower precision in the joint estimation of the parameters.
In predictive microbiology, the estimation of the parameters of the secondary model is carried out by fitting the values obtained from the estimation of the parameter $\mu_{max}$ in the previous step for each temperature. 
This fitting assumes the estimates of a parameter to be observed values, which is associated with an estimation error that is not considered. This paper proposes the use of the $c$-optimal design to obtain an estimate with minimum variance of the parameter $\mu_{max}$. 
In this way the fitting is made on a value with lower associated uncertainty. 
The results of the $c$-efficiency (Table \ref{tab:c-opt mumax}) of these designs compared to those used in practice show a reduction of more than 50\% in the number of required observations. \\

Now, in order to provide fair comparisons, the temperatures at which these experiments are designed are those taken from the study with which the results are compared. However, OED allows for the precise estimation of all parameters in a direct manner by designing in two variables, that is, providing not only the times but also the temperatures at which the experiment should be carried out. If the kinetic parameter ($\mu_{max}$) is replaced in the primary model by the secondary model, it provides a extended model in two variables, time and temperature, with all the parameters of interest. 
The $D$-optimal design in two variables, $\xi^\star_T$, 
results in a 42\% reduction in the number of observations with respect to the designs used in practice. 
Since this optimal design depends on the choice of initial values for the parameters, a sensitivity analysis of the design is carried out against uncertainty in these values. This analysis shows that the design is robust for all parameters except $\lambda$. 
To make the design more robust, the methodology developed by \cite{del2024new} could be applied, in which points are added to the support of the optimal design to minimise its loss of efficiency with respect to the uncertainty in the selected parameter(s). 
This methodology is based on the idea of the maximin criterion with a previous step, which guarantees computational savings, in which the candidate points to enter the design are preselected.\\

Two-variable $D$-optimal design leaves the experimenter to choose two of the observation points, the initial and final points. For the initial point (0h), the experimenter can choose the temperature at which the sample is taken. For the final point, due to the construction of the design, it can be observed that there is a relationship between the temperature at which the experiment is carried out and its final experimental time.
In order to save costs in the experiments, different reductions in the final experimental time of the experiment are proposed (Figure \ref{fig:curvas ahorro}), ensuring a minimum efficiency of the resulting design. 
These reductions show a minimum saving of 48\% in time with respect to the optimal time, with minimal losses in efficiency (between 1\% and 5\%). In addition, these relationships between temperature and final experimental time are expressed mathematically for various efficiencies (Table \ref{tab:relaciones_temp_tiempo}). This is of interest because the experimenter, knowing the relationship of costs of the experiment relative to time or temperature, can choose between waiting longer and experimenting at a lower temperature, or shortening the time of the experiment by increasing the temperature of the final observation.\\

Finally, in the case where the goal of the study is only to estimate the parameters of the secondary model, the $c$-optimal design of both parameters in two variables is proposed, $\xi^\star_{b,T_{min}}$. The $c$-efficiency of the previous $D$-optimal design, $\xi^\star_T$, with respect to this one is 1.5\%, that is, when the only concern is to estimate these two parameters, $D$-optimal design is not very efficient and a design intended solely for this estimation is necessary.\\

\section{Acknowledgements}

This work was supported by Ministerio de Ciencia e Innovación [grant number PID2020-113443RB-C21], by Junta de Comunidades de Castilla-La Mancha [grant number SBPLY/21/180501/000126] and by Universidad de Castilla-La Mancha, Plan Propio co-funded by FEDER [grant number 2022-GRIN-34288].

\bibliographystyle{apalike}
\bibliography{bibliografia}

\end{document}